\documentclass[sigconf]{acmart}

\AtBeginDocument{%
  \providecommand\BibTeX{{%
    \normalfont B\kern-0.5em{\scshape i\kern-0.25em b}\kern-0.8em\TeX}}}

\copyrightyear{2019}
\acmYear{2019}

\acmConference[]{Under Review}{2020}{}
\acmYear{}
\copyrightyear{2019}

\usepackage{textalpha}
\usepackage{graphicx}
\usepackage[T1]{fontenc}
\usepackage{booktabs}
\graphicspath{ {./images/} }

\begin{document}

\title{Addressing Purchase-Impression Gap through a Sequential Re-ranker}

\author{Shubhangi Tandon}
\email{shtandon@ebay.com}
\author{Saratchandra Indrakanti}
\email{sindrakanti@ebay.com}
\author{Amit Jaiswal}
\email{amit.jaiswal@gmail.com}
\authornote{This work was done while author worked at eBay}
\author{Svetlana Strunjas}
\email{sstrunjas@ebay.com}
\author{Manojkumar Rangasamy Kannadasan}
\email{mkannadasan@ebay.com}
\affiliation{%
  \institution{Search Science, eBay Inc.}
  \city{San Jose}
  \state{California}
}

%
%
%

\renewcommand{\shortauthors}{Tandon and Indrakanti, et al.}

\begin{abstract}
Large scale eCommerce platforms such as eBay carry a wide variety of inventory and provide several buying choices to online shoppers. It is critical for eCommerce search engines to showcase in the top results the variety and selection of inventory available, specifically in the context of the various buying intents that may be associated with a search query. Search rankers are most commonly powered by learning-to-rank models which learn the preference between items during training. However, they are pointwise during inference, and score items independent of other items. Although the items placed at top of the results by such scoring functions may be independently optimal, they can be sub-optimal as a set. This may lead to a mismatch between the ideal distribution of items in the top results vs what is actually impressed. In this paper, we present methods to address the purchase-impression gap observed in top search results on eCommerce sites. We establish the ideal distribution of items based on historic shopping patterns. We then present a sequential reranker that methodically reranks top search results produced by a conventional pointwise scoring ranker. The reranker produces a reordered list by sequentially selecting candidates trading off between their independent relevance and potential to address the purchase-impression gap by utilizing specially constructed features that capture impression distribution of items already added to a reranked list. The sequential reranker enables addressing purchase-impression gap with respect to multiple item aspects. Early version of the reranker showed promising lifts in conversion and engagement metrics at eBay. Based on experiments on randomly sampled validation datasets, we observe that the reranking methodology presented produces around 10\% reduction in purchase-impression gap at an average for the top 20 results, while making improvements to conversion metrics.

\end{abstract}




\maketitle

\section{Introduction}
Search plays a critical role on eCommerce sites, empowering online shoppers to discover and purchase items. Large scale eCommerce platforms such as eBay carry a wide variety of inventory that match a search query, varying in terms of properties or aspects of items in the recall set. These aspects could be retail properties such as item condition, price, shipping options or product-specific characteristics such as features, model variants or specifications. Further, based on individual buying intents, different users may be interested in shopping specific aspects of items for the same search query. To accommodate various buying intents, as open marketplaces, it is critical that search results in modern eCommerce websites showcase the variety and selection of inventory available.
 
Ranking search results, in general, is focused on determining the ordering of documents based on their relative relevance to maximize their utility. Learning-to-rank models aim to learn the notion of relative document relevance during training using pairwise or listwise loss functions to learn a scoring function \cite{li2011short}. However, during inference, most models are limited to pointwise or univariate scoring functions where the documents are scored independent of other items in the set. Ecommerce search result ranking, further, is tailored towards surfacing highly relevant and sellable inventory to users. Unlike web search, where the goal is to address an informational need, eCommerce search is focused on empowering users to contrast and compare inventory towards completing a purchase. Typically, many eCommerce sites order items in the recall set by their likelihood of sale (\textit{sellability}), as predicted by a learning to rank model. Irrespective of the loss function used in training the model (pointwise, pairwise, or listwise), inferencing is done in a point-wise fashion assuming independent document relevance. The ranking order thus produced, specifically in the top results that tend to have the largest influence on shoppers, may not be aligned with the objective of showcasing the selection of inventory available. Further, the rankers could deem items with certain aspects more sellable than other aspects due to observed majority buying behavior, which may lead to an over representation of the former in the top ranked results. While independently, each of the items in the top results may be more likely to be sold relative to others, they may be sub-optimal when perceived as a set of top results, specifically in the context of addressing various buying intents. This may lead to a gap between what shoppers expect to see in the top results of a search result page, compared to what is actually shown (\textit{impressions}). To that end, it is critical to methodically determine a suitable set of top items to be impressed to shoppers against a given search query.
 
To further illustrate the problem being addressed, we will use an example query \textit{laptop} in the context of item aspect \textit{condition}. In an over-simplified scenario for the purpose of illustration, based on the majority of historic sales, let’s say laptops in refurbished condition are more likely to be sold, assuming all other factors are equal. In this case, a pointwise scoring ranker may favor refurbished laptops, since, independently refurbished laptops are more likely to be sold. However, historic sales may also constitute considerable sales of new and used condition laptops. Although refurbished items may seem more suitable to be shown at the top of the results based on independent likelihood of sale, several minority buying intents may have been left unattended to in this scenario. Further, considering a set of top results as perceived by a user, we may be better off impressing certain new and used items as well, in order to showcase the selection of inventory and choice. This raises questions on what the appropriate distribution of condition aspect in top laptop results is, and how such a distribution can be enforced while ranking. Since a point-wise scoring ranker does not explicitly influence the distribution of items with respect to condition in this example, we must develop methods that facilitate the same.To summarize, the point-wise nature of most learning may lead to a mismatch between item aspect distribution of what shoppers purchase on average with respect to a query (\textit{desired distribution}) versus distribution of items actually impressed on a search results page (\textit{purchase-impression gap}). This brings up two primary questions:
\begin{enumerate} 
\item How do we determine the desired distribution of aspects for items against a given search query?
\item If we know the desired distribution, how do we enforce it in top results on a search results page?
 \end{enumerate}
Purchase behavior of shoppers over a period of time against a query can serve as a reasonable proxy for the desired distribution of aspects to be impressed. Through the application of navigational features and query reformulations, shoppers eventually buy what they intend to. Although the available inventory at a given instance and the behavior of the point-wise scoring ranker influences the distribution of item aspects impressed on a search results page, distribution of aspects with respect to items purchased by shoppers after issuing a query, can provide a global view of shopper preferences. In this work, we determine the desired distribution of aspects for each query from shopper purchase patterns observed over a period of time.
 
Once a desired distribution of aspects is established, the ranker must be aware of the other items being placed in the ranked results in order to be able to enforce the distribution. An active area of research, groupwise scoring functions learn to score a fixed-size group of items \cite{ai2019learning}. However, implementing groupwise scorers in the production environment of large scale eCommerce marketplaces is challenging, specifically in the context of computational cost and latency. Another approach that has been studied and implemented, especially in the context of diversification of search results, is sequential reranking of top results\cite{agrawal2009diversifying} \cite{zhu2014learning}. Sequential reranking is a greedy approach to reranking results by placing items in a ranked list sequentially from top to bottom, selecting the next candidate to be placed by taking into account the ones that have already been placed in the re-ranked list. To enforce the desired distribution, we implement a sequential reranker of top k search results produced by a conventional pointwise scoring ranker (henceforth referred to as textit{production ranker}). The aim of the reranker is to select the next candidate item to be appended to the re-ranked list such that the selection minimizes purchase-impression gap, based on items already added, while ensuring that the selected item is independently sellable. In other words, the reranker trades-off between a candidate item’s \textit{best-match} score produced by production ranker, and the candidate’s potential to minimize the purchase-impression gap on the search results page. This is achieved by modeling the perceived purchase-impression gap at each position using specially constructed features, referred to as aspect-impression-share features (\textit{ais} features), that take into account the aspects of items placed in higher positions, in conjunction with best-match score. 

In this paper, we present methods developed to address the purchase-impression gap that may be observed on eCommerce sites as a consequence of pointwise scoring functions used by most learning-to-rank models. We obtain a desired distribution of item aspects that should be impressed in the top results for a query by mining historical item aspect level purchase behaviors corresponding to the query. We then present a sequential reranker that reorders the top results from a conventional pointwise ranker, minimizing purchase-impression gap whilst selecting independently sellable items by employing \textit{ais} features. Early versions of the reranker with a small set of aspects launched on eBay search showed promising item conversion and user engagement improvements. Offline experiments on randomly sampled search behavior datasets show around 10\% reduction in purchase-impression gap while leading to improvements in conversion metrics such as sale rank and mean reciprocal rank of purchased items. The proposed methodology, on site implementation of the reranker and details on experiments and results are presented in the section that follow. 

\section{Related Work}
In eCommerce search, users either have a specific intent/product in mind or they are issuing a broad query to understand the breadth of inventory available. In either case, they end up performing a lot of comparison shopping and identifying tradeoffs that work for them before making a purchase decision. Users perform a joint evaluation of all the candidate products either on the search page or through the Cart or by Saving items to their wish lists. We will review some of the related work which focuses on how users evaluate items when multiple options are presented and how diversifying search results satisfies different user needs and helps them make purchase decisions faster. 
Lichtenstein et. al  \cite{lichtenstein1971reversals} presented some early work where user decisions are different when choices are presented separately compared to when they are presented together. Similarly, there are a number of previous studies analyzing the click behavior on a document and its influence on both rank and other documents in the presentation \cite{joachims2017accurately}\cite{joachims2007evaluating}\cite{craswell2008experimental}. We also measured and validated the influence of neighborhood on the preference of an item in ecommerce in our earlier work \cite{saratch2019influence}.
Several studies were also performed to showcase diverse search results given that there are different user needs for certain queries. They can be mainly divided into implicit and explicit diversification. Implicit approaches model similarity between documents and try to place documents which are similar to the query and dissimilar to the previously placed documents \cite{carbonell1998use}. Explicit approaches models certain aspects of the query like taxonomy \cite{agrawal2009diversifying}, query reformulations \cite{radlinski2006improving}, external resources \cite{he2012combining} etc and try to place documents that cater to all those aspects. All of these approaches apply heuristics based utility functions whereas Zhu et al addressed search results diversification as a learning problem where a ranking function is learned for diverse ranking of search results \cite{zhu2014learning}. Our work extends these previous approaches to adapt to eCommerce and focuses on effectively modeling different aspects of the query that are correlated with purchases. We also change the problem optimizing for diversity to showcasing the breadth of inventory that satisfies a purchase distribution measured on user behavior.

\section{Methodology}
Our aim is to close the gap between purchase and impression distributions measured based on conversion distribution of search results for a query and the recall distribution of the top $k$ items, where $k$ is a non-zero integer. We introduce a \textit{sequential reranker} on top of the existing \textit{production ranker}. It reranks top $k$ items to minimize the purchase-impression gap with respect to aspect and features constructed to capture impression share and historically mined purchase shares. 
Due to the nature of the sequential ranker, during online inferencing, after $m$ items have been ranked , we chose the $m+1^{th}$ item so that it has a good best match score and it also aligns with the purchase share distribution. The sequential reranker calculates a \textit{bridge\_score}  which is defined as :
	\[bridge\_score =  \sum w_{i}x_{i}  ,\]
	where $x_i$  is the feature and $w_i$ is the weight associated with the feature for the given query

The intermediate \textit{bridge\_score}  is weight-summed with the best-match ranking score from the production ranker to generate the final ranking score. The \textit{bridge\_score}  is weighed by the parameter  (1 -  $\alpha$ )/ $ \alpha$  where  $\alpha$ is a learned parameter that determines the relative importance of the two intermediate scores. For our current model , $\alpha$ is learned globally but we will be exploring query-wise $\alpha$ parameters in the future.
 \[ final\_score =   best\_match\_score + \frac{(1 -  \alpha )}{ \alpha }* bridge\_score \]
We build query-wise models, so for each query the weights are calculated separately.

\subsection{GMV Shares as weights}
The weights for the \textit{bridge\_score} are the aspect-specific Gross Merchandise Volume (GMV) shares that have been calculated by analysing historical behavioral data for a query. For the items that give a converting signal, we run a data pipeline job that records the values for certain aspects painted on the item (eg \textit{condition}, \textit{buying format} etc). We consolidate this data over a period of time to get the GMV share for each of this aspect as a distribution on its aspect values, summing up to 1. For instance, for the query \textit{iphone} if we observe that users are more likely to interact with new items as opposed to refurbished  or old in the ratio 5:3:2 , then the GMV share distribution will be [0.5, 0.3, 0.2] for new , refurbished and old aspect values respectively. 
Since the purchase distribution is different for each query, the weights are also differ by query. 

\subsection{Aspect Impression Share features}
Each aspect value for which we have a GMV share is a candidate Aspect-Impression-Share(\textit{ais}) feature . The \textit{ais} features represent how much a given feature differs from the distribution so far. It is greater than 0 if the feature (or aspect value ) is painted on the current item, 0 otherwise. 

For instance,  aspect \textit{condition} could have 3 \textit{ais} features: \textit{ais\_is\_new, ais\_is\_refurbished, ais\_is\_old}. To construct \textit{ais} features , we first generate binary features : \textit{is\_new, is\_refurbished ,is\_old } indicating the condition of the item . For example, if the item belongs to new condition, $is\_new =1$, $is\_refurbished = 0$ , $is\_old =0$.
Now we define an intermediate delta feature as described in \cite{indrakanti2019exploring}, which will capture how diverse the current feature is from the distribution that we have seen before. The following example illustrates \textit{ais} feature computation for a sample scenario. Let us assume we have already ranked 10 items and are selecting the $11^{th}$ to be appended, with 6 of the 10 items being new, 3  old and 1 in refurbished condition.. If the current item is new, the intermediate \textit{delta} features are: 
\[[delta\_is\_new, delta\_is\_old, delta\_is\_refurbished] = [0.4, 0.7, 0.9]\]
The \textit{ais} features are defined as : $delta\_feature * is\_feature $. Therefore the features can be computed to be  
\[ [ ais\_is\_new, ais\_is\_refurbished, ais\_is\_old] = [0.4, 0 ,0] \]

\section{Onsite Implementation}
The Sequential Ranker is implemented in production and re-ranks top $k$ results out of $m$ results  that are already produced and ranked by our Production Ranker. This ranker sequentially reorders results from the Production Ranker by starting from the highest ranked position and moving lower down the list, each time selecting the next candidate that maximizes the score defined by the re-ranking formula that is described in a previous section.

 \begin{figure}[h]
\includegraphics[width=8cm]{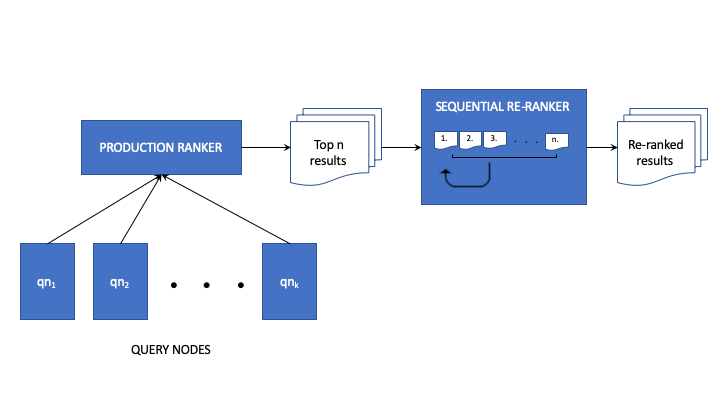}
\caption{Proposed Architecture with Sequential Reranker }
\label{fig:arch}
\end{figure}

Parameters for the sequential  re-ranker could be easily changed and customized  via a corresponding configuration file called a profile. Via a given profile, we can choose how many top results we want to reorder (previously mentioned value $k$) as well as how many results we want to consider when choosing item candidates (previously mentioned value $m$). The only prerequisite is that $k < m$. Besides choosing $k$ and $m$, we could also customize the parameter for sets of similar queries or even individual queries, which we plan to explore in the future,  and apply different re-ranking formulas for different queries. Figure \ref{fig:arch} provides an overview of the sequential reranking architecture.

\section{Results}

The metric that we want to minimize with the  reranker is the gap between the impression of top $k$ items and the learned gmv share distribution of a query. The purchase-gap is formally defined as the mean non-negative difference between a query’s expected GMV share and the impressed recall distribution across all its aspect values.  Note that, a gap only contributes if its greater than 0. We don’t penalize an aspect value to have more than the expected GMV share . 

\[Gap = 1/k *  \sum_{i=0}^{k} min((gmv\_share - (1 - ais\_feature)), 0)\]

We also ensure that the conversion metrics like mean reciprocal rank (MRR) are not impacted adversely in the process of re-ranking. 

To size up our problem, we randomly sampled 3000 queries to look at the purchase-impression gap. We saw that on average, there is a 13\% gap between sales and impression for each aspect (like condition) in this dataset for each query. Around 77\% of the queries had a Purchase-Impression Gap for the aspects we considered. 

We also conducted an A/B test where we would determine the Aspect with the highest Purchase-Impression Gap for a query and then boost items which aligned with the GMV Share distribution for the Aspect with the highest Gap. The Test did well and we saw significant lifts in Gross Merchandise Value Bought metric, along with other metrics. This feature has now been integrated in our Production Ranker.

For this experiment, we collected behavioral data for a time period of 14 days and collected over 80,000 Query Search Result Sessions upto top 50 Items for each query. We obtained the GMV Share by mining the historical Purchase Distribution for these queries . We ran the Sequential Ranker on top of the Production Ranker to rerank top 20 items from the top 50 items available. We trained with four different values of $\alpha$ : 1 ( which resolves to Best Match Score from Production Ranker), 0.8 , 0.5, 0.2. The results are summarized in table \ref{tab:my-table} . We see the biggest reduction in Gap when $\alpha$ = 0.2 so the Bridge Score has a much higher weight . However the MRR lift is very small. Therefore for our A/B test we choose to use $\alpha$ = 0.5 which reduces the Gap by 8\% and also gives a MRR Lift of 3.7\%. In the future, we will try to learn $\alpha$ at a query level instead of a query-set level. 

\begin{table}[]
\begin{tabular}{@{}llll@{}}
\toprule
$\alpha$ & \begin{tabular}[c]{@{}l@{}}Average Gap \\ Difference\end{tabular} & \begin{tabular}[c]{@{}l@{}}Delta MRR \\ Shift\end{tabular} & \begin{tabular}[c]{@{}l@{}}P-Value \\ for MRR\end{tabular} \\ \midrule
1.0        & 0                                                                 & 0                                                          & 0.9                                                        \\
0.8        & 2.3\%                                                             & 4.1\%                                                        & 0.6                                                        \\
0.5        & 8\%                                                               & 3.7\%                                                       & 0.5                                                        \\
0.2        & 15\%                                                              & 1.06\%                                                      & 0.5                                                        \\ \bottomrule
\end{tabular}
\caption{Avergage Gap Difference and MRR shift for different values of $\alpha$}
\label{tab:my-table}
\end{table}

 \begin{figure}[h]
\includegraphics[width=8cm]{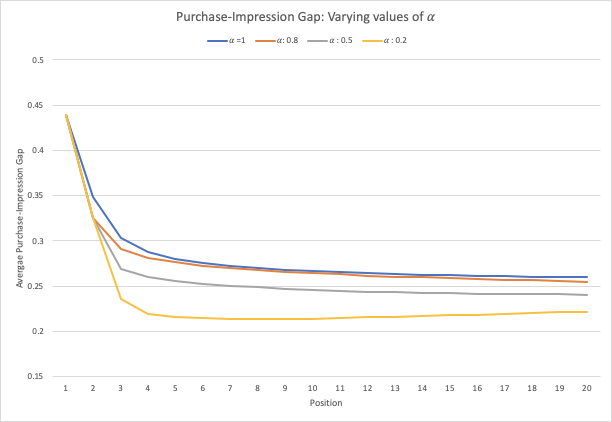}
\caption{Average Position-wise Purchase-Impression Gap for Queryset }
\end{figure}

\section{Conclusion}
Most commonly used learning-to-rank models score items independent of others while ranking, irrespective of pairwise or listwise loss functions used during training. Search rankers on eCommerce sites such as eBay generally employ rankers trained on sellability of items against a query. However, the pointwise nature of scoring can lead to an over-representation of items with certain aspects over others in the top search  results, leading to a purchase-impression gap, i.e. a mismatch between what shoppers purchase on average with respect to a query versus what is actually shown. The effects of this are even more pronounced on eCommerce sites where shoppers compare and contrast items to make buying decisions, as compared to web search. To address such a mismatch, we must develop ranking methods that are aware of the other items being placed in a ranked search result page. Further, such methods must be able to enforce a desired distribution of aspects in the top results. To that end, we developed methods to establish a preferred distribution of aspects to be impressed in the top search results against a query. We implemented a sequential reranker that reorders the top results produced by pointwise scoring production ranker. The sequential reranker manages a tradeoff between best-match score which represents independent sellability of an item and bridge-score, the item’s potential to minimize purchase-impression gap. 

We mine historical buying patterns with respect to individual search queries to establish a preferred aspect distribution of item aspects with respect to each query. We then apply the linear sequential reranker to rerank top best-match results from production ranker. Experiments on randomly sampled validation datasets indicate that the presented methods lead to a significant reduction in average purchase-impression gap measured over the top 20 reranked results. Early versions of this implementation based on a small selected set of item aspects launched on eBay search sites lead to statistically significant lifts in conversion and engagement metrics on search result pages.

While the sequential reranker presented in this work, for the purpose of simplicity, is a linear model that manages a straightforward tradeoff between best-match score and bridge score, it can be easily extended to more complex models. For instance, a deep neural network can be trained to learn this tradeoff as illustrated in \cite{zhu2014learning}. A smaller set of popular global item aspects were used in this work. The set of aspects can be extended to a larger set that includes local item or query-specific aspects to improve the reranker. Further, We employ separate features to represent individual item aspects. The framework can be extended to consume embeddings for items learnt over an item aspect space to facilitate bridge score computation. In summary, we presented an intuitive approach to address an important problem observed on large eCommerce sites such as eBay. We employ insights from historic purchase behavior to debias existing ranking through a simple yet powerful and extensible framework that provides a scalable solution without adding significant latency to site experiences.

\bibliographystyle{ACM-Reference-Format}
\bibliography{acmart}

\appendix

%
%
%
%
%
%
%

\end{document}